\documentstyle[preprint,aps,prb]{revtex}
\begin{document}

\draft
\title{The resistance anomaly in the surface layer
of Bi$_2$Sr$_2$CaCu$_2$O$_{8+x}$ single crystals under
radio-frequency irradiation}

\author{Nam Kim, Hyun-Sik Chang, Hu-Jong Lee, and Yong-Joo Doh}

\address{Department of Physics,
Pohang University of Science and Technology \\
Pohang 790-784, Republic of Korea}

\maketitle

\begin{abstract}
We observed that radio-frequency (rf) irradiation significantly
enhances the $c$-axis resistance near and below the
superconducting transition of the CuO$_{2}$ layer in contact with
a normal-metal electrode on the surface of
Bi$_{2}$Sr$_{2}$CaCu$_{2}$O$_{8+x}$ single crystals. We attribute
the resistance anomaly to the rf-induced charge-imbalance
nonequilibrium effect in the surface CuO$_{2}$ layer. The
relaxation of the charge-imbalance in this highly anisotropic
system is impeded by the slow quasiparticle recombination rate,
which results in the observed excessive resistance.

\end{abstract}

\newpage
\tightenlines \narrowtext

\section{INTRODUCTION}
Recently, the resistance anomaly (RA) characterized by a
pronounced resistance enhancement above the normal-state value
$R_N$ near the superconducting transition has been observed in
various systems such as superconducting aluminum wires
\cite{Strunk98,Strunk96,Ju-Jin} and structures including
artificial normal-metal/superconductor (NS) interfaces.\cite{Park}
Most of the previous studies on this novel phenomenon indicate
that the resistance enhancement takes place either in
quasi-one-dimensional structures\cite{Strunk98,Strunk96,Ju-Jin} or
in thin films\cite{Park} in their superconducting state when the
voltage is probed within the range comparable to the
characteristic length scales governing the superconducting
transition. More recently, it has been found that radio frequency
(rf) irradiation controls the appearance and the magnitude of the
RA. This leads to explanations for the RA in terms of
nonequilibrium superconductivity, such as charge-imbalance
phenomenon arising either from rf-excited phase-slip centers or
artificial NS interfaces, mixing of extrinsic rf noise, etc.

In this study we have observed similar resistance enhancement in
$c$-axis conduction measurements on small (junction area of $25
\times 35$ $\mu$m$^{2}$) stacks of Bi$_2$Sr$_2$CaCu$_2$O$_{8+x}$
(Bi2212) single crystals irradiated by rf waves, near or below the
dc superconducting transition. The RA is observed only for a
three-probe measurement configuration, where the voltage
measurement includes the surface CuO$_2$ layer which is in contact
with a normal-metal (Au) electrode. The RA in this system
resembles the one observed in the conventional aluminum
wires\cite{Strunk96} in that an excess resistance occurs near or
below the superconducting transition and current-voltage ($IV$)
characteristics exhibit a nonlinear excess voltage at zero bias.
Since the anomaly occurs in the $c$-axis resistance of a structure
including a NS interface (in this case, the interface consists of
a Au electrode and the surface CuO$_2$ layer of Bi2212 single
crystals) we believe the RA is related to the charge-imbalance
nonequilibrium state produced by rf-induced ``hot" quasiparticles
injected from the normal-metal region into the superconducting
region.\cite{Strunk98} We assume that the rf irradiation excites
quasiparticles to higher energy states above the gap of the
surface superconducting layer. The enhanced accumulation of
charge-imbalanced quasiparticles in the surface CuO$_2$ layer upon
rf irradiation increases the chemical potential of quasiparticles
in the layer, which can lead to the enhancement of the measured
voltage across the NS interface. The charge-imbalanced
nonequilibrium state is known to relax to charge-balanced state
within the characteristic time $\tau_{Q^*}=(4/\pi) \tau _{i}[k_B
T/\Delta (T)]$ at temperatures near the superconducting
transition, where $\tau _{i}$ is the inelastic scattering time,
and $\Delta (T)$ the energy gap.\cite{Clark} The resulting excess
resistance due to the charge imbalance becomes most prominent near
the superconducting transition, because the relaxation becomes
significantly slower as the fraction of the effective states
participating in the relaxation, $\Delta/k_B T$, is reduced near
the transition.\cite{Tinkham2}

To our knowledge, no experimental observations of charge-imbalance
phenomena of this kind have been reported before in high-$T_c$
superconductors (HTSC). This lack of the observation of the
phenomena may result from the high phonon-induced inelastic
scattering rate near the superconducting transition of HTSC. The
electron-phonon scattering rate $\tau_{e-ph} ^{-1}( \approx
\tau_{i}^{-1})$, depending on the quadratic power of the
temperature in the two-dimensional clean limit which may be
relevant to the surface conduction layer, can be a few hundred
times larger in HTSC than in conventional superconductors near the
transition. Thus, the probability of HTSC being in a
charge-imbalance nonequilibrium state is reduced as much and to
date no RA has been reported for the materials. As we reported
previously,\cite {Kim} however, the $d$-wave superconductivity of
the surface CuO$_{2}$ conduction layer in Bi2212 single crystals
in contact with a normal-metal electrode is suppressed due to the
proximity effect. The superconducting transition temperature of
the surface layer, $T_{c}^{'}$, is reduced accordingly to at least
about one-third of the transition temperature of the inner CuO$_2$
layers, $T_{c}$. The resultant reduced electron-phonon scattering
rate near $T_{c}^{'}$ in the surface CuO$_{2}$ layer enhances the
possibility of observing the nonequilibrium RA effect under
rf-wave irradiation.

\section{EXPERIMENT}
Bi2212 single crystals were grown by the standard
solid-state-reaction method.\cite {Chung} Upon cleaving a Bi2212
single crystal a 50 nm-thick layer of Au was thermally deposited
on the top of the crystal to protect the surface from
contamination during the further fabrication process as well as to
obtain a clean interface between the normal-metal electrodes and
the Bi2212 single crystal. Stacks with junction area of $25 \times
35$ $\mu$m$^{2}$ and the thickness of about $15$ nm were then
patterned using the conventional photolithography and Ar-ion-beam
etching. A stack of that thickness contains about 10 CuO$_2$
layers. Further details of the sample fabrication are described
elsewhere.\cite{Kim}

The temperature dependence of the $c$-axis resistance $R_c (T)$ of
a stack was measured by the conventional lock-in technique and
$IV$ data by a dc method. The schematic measurement configuration
is shown in the inset of Fig. 1(a). All measurements were done in
a three-terminal configuration with a low pass filter being
connected to each measurement electrode. Three-terminal
configuration was adopted to probe the surface junction by
intentionally including the potential drop across the surface
layer. The contact resistance between the Au electrode and the
surface CuO$_{2}$ layer was less than 0.1 $\Omega$.\cite{Doh1} An
rf wave with frequency of 70 MHz was transmitted through coaxial
cables to the current leads as shown schematically in the inset of
Fig. 1(a). A capacitor was inserted between the specimen and the
rf generator to separate the rf signal from the dc probing
current. Because of the uncertainty in the rf coupling we were not
able to determine the actual rf power transmitted to the specimen.
Therefore the power levels specified below are the nominal values.

\section{RESULTS AND DISCUSSION}
Fig. 1(a) shows the typical $c$-axis resistance of a stack of
intrinsic junctions, $R_c (T)$, in the three-probe configuration
with the rf irradiation being turned on (open triangle) or off
(open circle). The dc bias current was 500 nA, which corresponds
to almost zero bias in the $IV$ curves shown in Figs. 2(a) and
2(b). The frequency and the power of the rf wave were 70 MHz and
-20 dBm, respectively. In the absence of the rf irradiation it is
known that the $R_c (T)$ for a stack on the Bi2212 single crystal
as illustrated in the inset of Fig. 1(a) exhibits a double
superconducting transition; the upper transition at $T_{c}$
($\simeq90$ K) is the one for the inner CuO$_{2}$ layers and the
lower one at $T_{c}^{'}$ is for the top-most CuO$_{2}$ layer in
contact with the Au electrode. The superconductivity in the thin
CuO$_2$ surface layer is suppressed by the proximity contact to
the thick normal-metallic Au electrode.\cite{Kim} Although all the
inner layers are Josephson coupled in the temperature range of
$T_{c}^{'}<T<T_{c}$, $R_c (T)$ remains finite because the surface
layer is in the normal state. The value of $T_{c}^{'}$ varies from
sample to sample. It was 35 K for our specimen in this study.

When an rf wave is irradiated the $R_c (T)$ curve deviates below
$T_{c}^{'}$ from the one in the absence of rf irradiation. Instead
of dropping, the $R_c (T)$ keeps increasing rapidly below
$T_{c}^{'}$, before eventually showing a superconducting
transition at a temperature further below $T_{c}^{'}$. The
superconducting transition for a given rf power is much sharper
than in the absence of rf irradiation. The $R_c (T)$ curve above
$T_{c}^{'}$ is, however, insensitive to the rf irradiation, which
indicates that the excess resistance is caused by the surface
CuO$_{2}$ layer. As the rf power increases the superconducting
transition temperature of the surface CuO$_{2}$ layer decreases,
along with an increase in the peak value of the RA as shown in
Fig. 1(b). For rf power of -20 dBm the RA becomes about 7 times
higher than the normal-state resistance at 100 K, which cannot be
explained by local heating due to rf irradiation. Since the rf
frequency used in this experiment was far lower\cite{Doh2} than
the plasma frequency $\omega_{p}$ the rf irradiation may have
acted as a dc bias and suppressed the transition
temperature.\cite{Tinkham} It is also possible that the rf-induced
pair breaking across the significantly reduced superconducting gap
near $T_{c}^{'}$ suppressed the superconducting transition.

We also plotted (dotted line) in Fig. 1(a) the expected $R_c (T)$
at temperatures below $T_{c}^{'}$ for the
normal-metal/insulator/$d$-wave superconductor (NID) junction
formed by the surface CuO$_{2}$ layer in its normal state and the
neighboring inner superconducting CuO$_{2}$ layer with $d$-wave
symmetry.\cite{Kim} A noticeable feature is that the observed
resistance enhancement near and below the superconducting
transition is far above the estimated tunneling resistance for a
NID-junction configuration. The RA, thus, cannot be simply
explained by the equilibrium tunneling consideration.

Fig. 2(a) shows the $IV$ characteristics at various temperatures
below $T_c^{'}$ when the rf irradiation of frequency 70 MHz and
power -20 dBm is on (solid line) or off (dotted line). For 29.8 K
data the dotted line is not discernible because of the high
overlap of the two sets of curves. Both the $IV$ curves at $T=4.2$
K and $T=29.8$ K show DID and NID tunneling characteristics,
respectively,\cite{Kim} regardless of the rf irradiation. On the
other hand, near the superconducting transition (14.9 K - 23.9 K)
the $IV$ curves in the presence of the rf irradiation show
multiple-transition features. For instance, the $IV$ curve at 14.9
K shows a double transition, one below and the other above the
critical current $I_{c}^{'}$ of the intrinsic surface junction in
the absence of rf irradiation.

In Figure 2(b) we show a close up view of the zero-bias region of
the $IV$ characteristics for the three temperatures near the
superconducting transition. The $IV$ characteristics at $T=18.9$ K
and $T=23.9$ K exhibit excess resistances above the normal-state
resistance near the zero bias, which correspond to the RA in the
$R_c (T)$ near the transition. A resistance enhancement due to the
appearance of rf-induced phase-slip centers in only 0.3-nm-thick
conducting layer is inconceivable. Burk {\it et al.} has proposed
\cite{Burk} that RA can be generated when the high-frequency
signal is mixed with the low-frequency or dc measuring current for
samples exhibiting nonlinear $IV$ characteristics. Since $IV$
characteristics of our sample become nonlinear near $T_{c}$ of
inner junctions as well as near $T_{c}^{'}$ of the surface
junction, the model predicts the appearance of the RA near the two
temperatures. Since the RA appeared only below $T_{c}^{'}$ in our
sample, however, we exclude the possibility of external rf mixing,
too.

We attribute the RA to the charge imbalance induced in the
superconducting surface layer. The excess resistance $R_{exc}$ due
to the charge imbalance at an interface consisting of a normal
metal and a conventional s-wave superconductor is given by\cite
{Clark}
\begin{equation}
R_{exc}=\frac{Z(T) \tau_{Q^{*}}}{2e^{2}N(0)\Omega},
\end{equation}
where
\begin{equation}
Z(T)=2 \int_{\Delta}^{\infty}N_{s}(E) \left[- \frac{\partial
f}{\partial E} \right]dE,
\end{equation}
$N(0)$ and $\Omega$ are the quasiparticle density of states of S
electrode at the Fermi level in its normal state and the volume of
the S electrode, respectively. $\tau_{Q^*}=(4/\pi) \tau _{i}[k_B
T/\Delta (T)]$ is the charge-imbalance relaxation time introduced
above. It should be noted, however, that Eq. (1) is applicable to
a homogeneous superconducting electrode with its length much
longer than the charge-imbalance relaxation length
$\Lambda_{Q^{*}} [=\sqrt{D\tau_{Q^*}}\sim 50 \mu$m if we assume
the diffusivity $D$ to be\cite{Liu} 31 cm$^2$/sec].

We use Eq. (1) to estimate the excess resistance in our specimen,
assuming that the charge imbalance is confined in the
superconducting surface CuO$_{2}$ layer, i.e., $\Lambda_{Q^{*}}$
is set to be the thickness of the surface layer, 0.3 nm. The
assumption may be justified by the existence of a strong
scattering barrier between the surface and the inner CuO$_{2}$
layers. The almost full opening of the superconducting gap in the
inner CuO$_{2}$ layers in the temperature range near $T_c ^{'}$
also reduces the probability of quasiparticle injection into the
inner layers from the surface layer. For the normalized density of
states of S, $N_{s}(E)$, we use the angle-averaged expression for
the $d_{x^2-y^2}$ symmetry\cite{Won} as $N_{s}(E)=Re\left[
\frac{1}{2\pi} \int_{0}^{2\pi} \frac{E}{\sqrt{E^2-[\Delta(T)
\cos(2\theta)]^2}} d\theta \right]$ by assuming that the surface
CuO$_2$ layer also has the $d_{x^2-y^2}$-wave symmetry. Fig. 3
shows the best fit to the tail of the resistive transition as a
function of reduced temperature under rf irradiation of three
different power levels. In calculation we adopted
$\Delta(0)/k_{B}T_{c}^{'}=3.85$ as obtained previously\cite{Kim}
and assumed that $\Delta(T)$ follows the BCS gap equation. We also
set $N(0)$ to be\cite{Okazaki} 0.8 states/[eV][Cu-atom] and adopt
the nominal value for the effective volume of the superconducting
electrode, $\Omega$=25 $\mu$m$\times$35 $\mu$m$\times$0.3 nm. The
inelastic scattering time $\tau_{i}$ is taken as the fitting
parameter. The best fit values of $\tau_{i}$ depend on the rf
irradiation power and turn out to be $2.7 \times 10^{-8}$ sec,
$2.0 \times 10^{-8}$ sec, and $4.6 \times 10^{-9}$ sec for the
power of -20 dBm, -23 dBm, and -30 dBm, respectively. $\tau_{i}$
gets longer as the transition temperature decreases for higher rf
power. Fits below the superconducting transition with only one
fitting parameter $\tau_{i}$ turn out to be reasonably good.

The values of $\tau_{i}$ determined in this way appear
unreasonably long compared to the ones obtained from the $R(T)$
data above $T_{c}$ based on the two-dimensional superconducting
fluctuation theory. For instance, the authors of Ref. [16]
estimated $\tau_{i}$ to behave as
$\tau_{i}=0.75\times10^{-13}$(100 K/$T)^2$ sec for Bi2212
whiskers, which would give a few orders of magnitude shorter
inelastic scattering time in the temperature range used in this
experiment. As described below we believe this long inelastic
scattering time stems from the two impeding mechanisms, both
spatial and temporal, for effective charge-imbalance relaxation.
In general, a charge-imbalanced nonequilibrium state relaxes to a
charge-balanced nonequilibrium state, where pairs of conjugate
electron- and hole-like quasiparticles relax to the paired ground
state within the quasiparticle recombination time, $\tau_{rec}$,
which is essentially given by\cite{Tinkham2} $\tau_{r}=3.7 \tau
_{i}[k_B T/\Delta (T)]$. In addition to this temporal relaxation,
in conventional quasi-one-dimensional systems including a NS
interface with homogeneous S electrode, any {\it local}
nonequilibrium state in S can diffuse away within the
characteristic length scale, $\Lambda_{Q^*}$, which is the spatial
relaxation. In our highly anisotropic Bi2212 system, however, the
spatial relaxation is hampered because the tunneling of
qusiparticles in the surface CuO$_{2}$ layer to the neighboring
inner CuO$_{2}$ layer is effectively blocked by the presence of
the finite gap in the inner layer near $T_c ^{'}$. In addition,
the recombination rate of the quasiparticles within the surface
CuO$_{2}$ layer becomes significantly slower as the gap shrinks as
it approaches $T_{c}^{'}$. As a result of these two mechanisms the
low-lying quasiparticle energy levels are highly occupied by the
accumulating quasiparticles, blocking the further relaxation of
the charge-imbalance nonequilibrium. Thus, in this situation, even
the charge-balanced state is in nonequilibrium and the relaxation
of the charge imbalance itself may slow down in the surface layer.
It significantly enhances the charge-imbalance-induced RA in the
surface layer and may explain the very long charge-imbalance
relaxation time we obtained. In this sense, the quasiparticle
recombination time is the bottleneck in the effective relaxation
of the charge-imbalance nonequilibrium in the surface layer. Thus,
one can conclude that what was actually measured in this
experiment was the quasiparticle recombination time. In fact, our
values from the fit above are in a reasonable range if we compare
them with the quasiparticle recombination time, 0.8 ns and 2
{$\mu$}s, obtained in stacks of Bi2212 single crystals by Tanabe
{\it et al.}\cite{Tanabe} and Yurgens {\it et al.}\cite{Yurgens},
respectively, from fits to the back-bending of $IV$
characteristics, another nonequilibrium effect in dc measurements.

Although we take into account the $d_{x^2-y^2}$-wave symmetry of
the superconducting state of the surface CuO$_2$ layer by adopting
angle-averaged quasiparticle density of states for the fit, the
effect of the existence of nodes in the gap is not directly taken
into account. No microscopic theory on the generation and the
relaxation of the charge-imbalance nonequilibrium state for a NS
interface where S has a $d_{x^2-y^2}$ order-parameter symmetry is
available to date. Naive consideration may lead to a conclusion
that the exsistence of nodes makes the generation of the charge
imbalance easier near the node region in the $k$ space. It may
also slow down the relaxation process near the node region at any
temperatures below the superconducting transition. Theoretical
studies are required on the systems involving superconductors with
$d_{x^2-y^2}$-wave symmetry.

\section{ACKNOWLEDGMENTS}
We appreciate the useful discussion with Prof. Soon-Gul Lee on the
charge-imbalance phenomena. This work was supported by BSRI, MARC,
KOSEF, and POSTECH.

\begin{figure}
\caption{(a) The $c$-axis resistance, $R_c (T)$, in the absence
(open circle) or in the presence (open triangle) of rf
irradiation. The frequency and the power of the rf wave was 70 MHz
and -20 dBm, respectively. The dotted line represents the
equilibrium tunneling resistance of the surface junction which is
assumed to be in a normal-metal/insulator/$d$-wave-superconductor
configuration. \cite{Kim} Inset: a schematic view of the sample
geometry and measurement configuration. (b) $R_c (T)$ curves with
varying rf-irradiation power, taken with the bias level of 500
nA.}
\end{figure}

\begin{figure}
\caption{ (a) $IV$ characteristics at various temperatures below
the superconducting transition at $T_{c}^{'}$ of the surface
CuO$_{2}$ layer, in the absence (dotted line) and in the presence
(dotted line) of the rf irradiation. The frequency and the power
of the rf wave was 70 MHz and -20 dBm, respectively. For clarity,
$IV$ curves for different temperatures are shifted horizontally.
(b) A close-up view of $IV$ characteristics near zero bias in the
presence of the rf irradiation at three different temperatures
near the superconducting transition as denoted by arrows in the
inset.}
\end{figure}

\begin{figure}
\caption{The $c$-axis resistance vs reduced temperature for
different rf power. The frequency of the rf wave was 70 MHz and
the values of the irradiation power for the curves were -30 dBm
(open circle), -23 dBm (open diamond), and -20 dBm (open
triangle). The corresponding values of $T_{c}^{'}$ for the three
curves were 26.3 K , 23.4 K, and 19.7 K, respectively. The solid
lines are the calculation results based on Eq. (1), for the
best-fit values of $\tau_{i}$ being $4.6 \times 10^{-9}$ sec, $2.0
\times 10^{-8}$ sec, and $2.7 \times 10^{-8}$ sec, respectively.}
\end{figure}


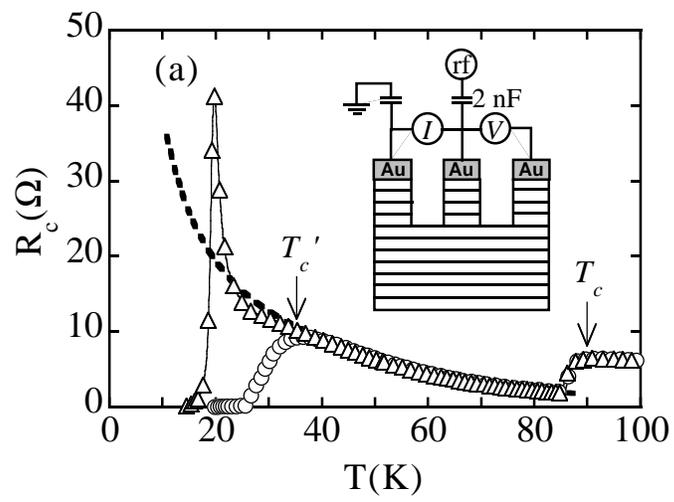
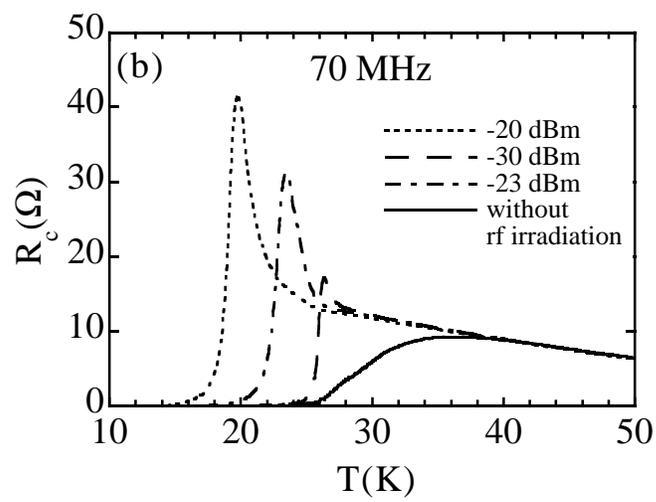
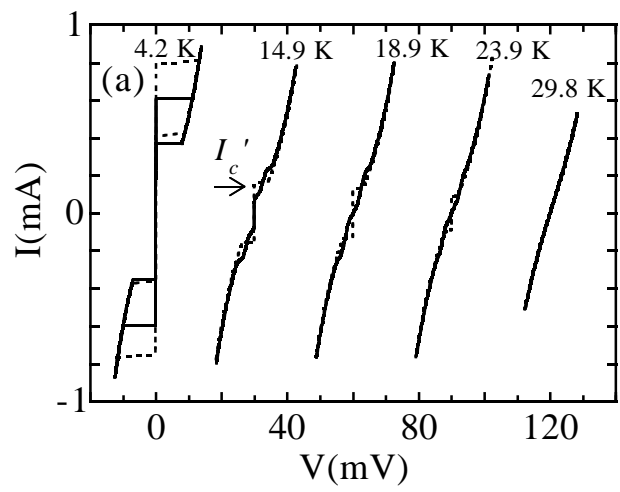
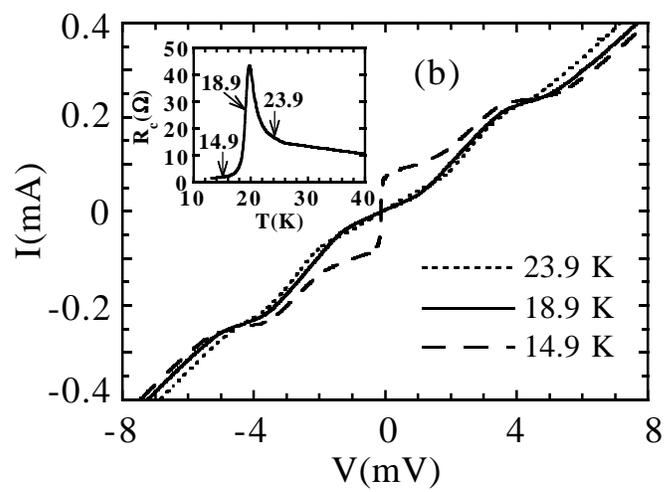
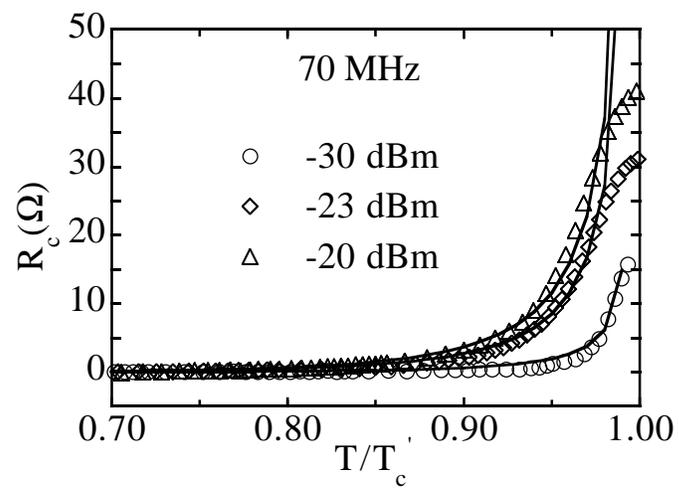
\end{document}